\documentclass[11pt]{article}
\usepackage{times}
\usepackage{geometry}
\usepackage{tabularx}
\usepackage[utf8]{inputenc}
\usepackage{enumitem,amssymb}
\usepackage{ragged2e}
\usepackage{graphicx}
\usepackage{comment}
\usepackage{multicol}
\usepackage[usenames]{xcolor} 
\definecolor{xlinkcolor}{cmyk}{1,1,0,0}
\usepackage{url}

\usepackage[
 colorlinks=true,    
 linkcolor=xlinkcolor,     
 citecolor=xlinkcolor,     
 filecolor=xlinkcolor,  
 urlcolor=xlinkcolor,      
 final=true
]{hyperref}

\usepackage[sort&compress,comma]{natbib}

\usepackage{enumitem}
\usepackage{wrapfig}
\usepackage{graphics}
\usepackage{graphicx}
\setenumerate{itemsep=0mm}

\begin{document}

\begin{raggedright} 
\huge
Snowmass2021 - White Paper \hfill \\[+1em]
\textit{Real-time Cosmology with High Precision Spectroscopy and Astrometry} \hfill \\[+1em]
\end{raggedright}

\normalsize

\noindent {\large \bf Thematic Areas:}  (check all that apply $\square$/$\blacksquare$)

\noindent $\square$ (CF1) Dark Matter: Particle Like \\
\noindent $\square$ (CF2) Dark Matter: Wavelike  \\ 
\noindent $\blacksquare$ (CF3) Dark Matter: Cosmic Probes  \\
\noindent $\blacksquare$ (CF4) Dark Energy and Cosmic Acceleration: The Modern Universe \\
\noindent $\square$ (CF5) Dark Energy and Cosmic Acceleration: Cosmic Dawn and Before \\
\noindent $\square$ (CF6) Dark Energy and Cosmic Acceleration: Complementarity of Probes and New Facilities \\
\noindent $\blacksquare$ (CF7) Cosmic Probes of Fundamental Physics \\
\noindent $\square$ (Other) {\it [Please specify frontier/topical group]} \\

\noindent{\large \bf Authors:}
Sukanya Chakrabarti$^{1}$, Anthony H. Gonzalez$^{2}$, Steve Eikenberry$^{3}$, David Erskine$^{4}$, Mustapha Ishak$^{5}$, Alex Kim$^{6}$, Eric Linder$^{6}$, Andrei Nomerotski$^{7}$, Michael Pierce$^{8}$, Anze Slosar$^{7}$, Paul Stankus$^{7}$, \mbox{Yu-Dai Tsai$^{9,10}$}\\
\\
$^1$ Rochester Institute of Technology\\
$^2$ University of Florida\\
$^3$ University of Central Florida\\
$^4$ Lawrence Livermore National Laboratory\\
$^5$ The University of Texas at Dallas\\
$^6$ Lawrence Berkeley National Laboratory\\
$^7$ Brookhaven National Laboratory\\
$^8$ University of Wyoming\\
$^9$ University of California, Irvine\\
$^{10}$ Fermilab\\

\clearpage

\noindent


\section{Introduction} 

Breakthroughs in physics and astrophysics are often driven by technological advances, with the recent detection of gravitational waves being one such example. This white paper focuses upon how improved astrometric and spectroscopic measurements from a new generation of precise, accurate, and stable astronomical instrumentation can address two of the fundamental mysteries of our time -- dark energy and dark matter -- and probe the nature of spacetime.

Instrumentation is now on the cusp of enabling new cosmological measurements based on redshifts (cosmic redshift drift) and extremely precise time-series measurements of accelerations, astrophysical source positions (astrometry), and angles (cosmic parallax). These allow tests of the fundamental framework of the universe (the Friedmann equations of general relativity and whether cosmic expansion is physically accelerating) and its contents (dark energy evolution and dark matter behavior), 
while also anchoring
the cosmic distance scale ($H_0$). 

The unexpected accelerated expansion of the Universe must arise from physics beyond the Standard Model: a dark energy or vacuum energy of a new field, or a breakdown in General Relativity. To date, this acceleration has been
inferred from the expansion history measured by distances through  cosmological probes including Type~Ia supernovae (SN), baryon acoustic oscillations (BAO), and the cosmic microwave background 
\citep[CMB; see][and references therein]{scolnic2018,planck2020,alam2021}.
Direct measurement of acceleration of the cosmic expansion would test the 
Friedmann-Lema{\^{\i}}tre-Robertson-Walker framework itself, and provide a new probe of cosmic expansion and dark energy. This has been a goal for over 60 years \citep{mcvittie1962,sandage1962}, and is finally within reach given technological developments that enable accurate measurement of the very small change in an object's redshift with observer time (the second derivative of position, i.e.\ the acceleration). Measurements of this redshift drift can reveal the physical nature of cosmic acceleration as well as parameter estimation  with precision competitive with and highly complementary to standard methods -- giving a Stage IV experiment the power of a Stage V one. 

This same technology can be applied to extremely precise time-series measurements of velocities to determine accelerations of sources within our Galaxy or nearby ones, creating a direct map of the gravitational field of the galaxy. The dark matter mass distribution, clustering, and any nonstandard interactions can be revealed through such maps. 
With current generation spectrographs like ESPRESSO \citep{Pepe2021} and NEID \citep{NEID_optical} expected to achieve radial velocity (RV) precision of $\sim$ 10 cm/s, one can directly measure the \emph{changes} in the RVs over decade baselines to obtain a line-of-sight acceleration.  These instruments thus far are on less than 10-m telescopes, and therefore cannot access the entire volume of our Galaxy and are limited to observations of relatively bright stars within a few kiloparsec from the Sun.  Future instruments on the Extremely Large Telescopes (ELTs) will be able to carry out direct acceleration measurements \emph{across} the Galaxy, and beyond. A key feature of such direct acceleration measurements is that the relative precision improves with time, such that decade-scale precision measurements of dark matter in the Galaxy are feasible if the instruments are designed to yield stable RV measurements on this timescale.  This technique thus far has largely been to detect and characterize exoplanets, but it is equally viable for understanding the nature of dark matter.

High precision positional and angle measurements (astrometry), 
leveraging large-aperture telescopes and extended time baselines, have the potential to enable direct measurements of secular parallax (seeing an object shift position on the sky due to its motion and cosmic expansion). Such geometric distances beyond our local group of galaxies would provide a new, more robust anchor for the cosmic distance scale. 
Quantum-assisted optical interferometers are one path for dramatic improvement of these astrometric measurements.

\section{Cosmological Redshift Drift }

Einstein's Equivalence Principle teaches us
that acceleration {\it is\/} gravitation and defines the curvature of spacetime. 
Cosmic acceleration -- the change in the expansion rate of the Universe -- is thus a fundamental tool for understanding the Universe and a signpost of a new realm of physics that directly addresses one of the key goals of DOE HEP and NSF PHY, ``exploring the basic nature of space and time''.

Cosmic acceleration is observable as a change in the measured redshifts of objects, known as cosmic redshift drift.  In 1962, McVittie laid out the relation of  redshift drift to spacetime acceleration, and Sandage proposed the use of the greatest facilities of the time to observe it \citep{mcvittie1962,sandage1962}. 
Redshift probes the spacetime as
\begin{equation}
1+z=\frac{(g_{\mu\nu}k^\mu u^\nu)_{\rm em}}{(g_{\mu\nu}k^\mu u^\nu)_{\rm obs}}\ \ .
\end{equation}
The cosmic redshift drift $dz/dt_{\rm obs}$ thus directly reveals
the evolution of the metric $g_{\mu\nu}$ (e.g.\ through Hubble expansion),
any interactions of the photon four-momentum $k^{\mu}$, and any
inhomogeneous accelerations -- evolution of the peculiar velocities $u^\nu$.
In the standard FLRW cosmology,
\begin{equation}
\frac{dz}{dt_{\rm obs}}=(1+z)\,H_0-H(z)\,, 
\end{equation} 
giving a redshift drift of $\mathcal{O}(10^{-10}(\Delta t/{\rm yr}))$. 

\begin{figure}
    \centering
    \includegraphics{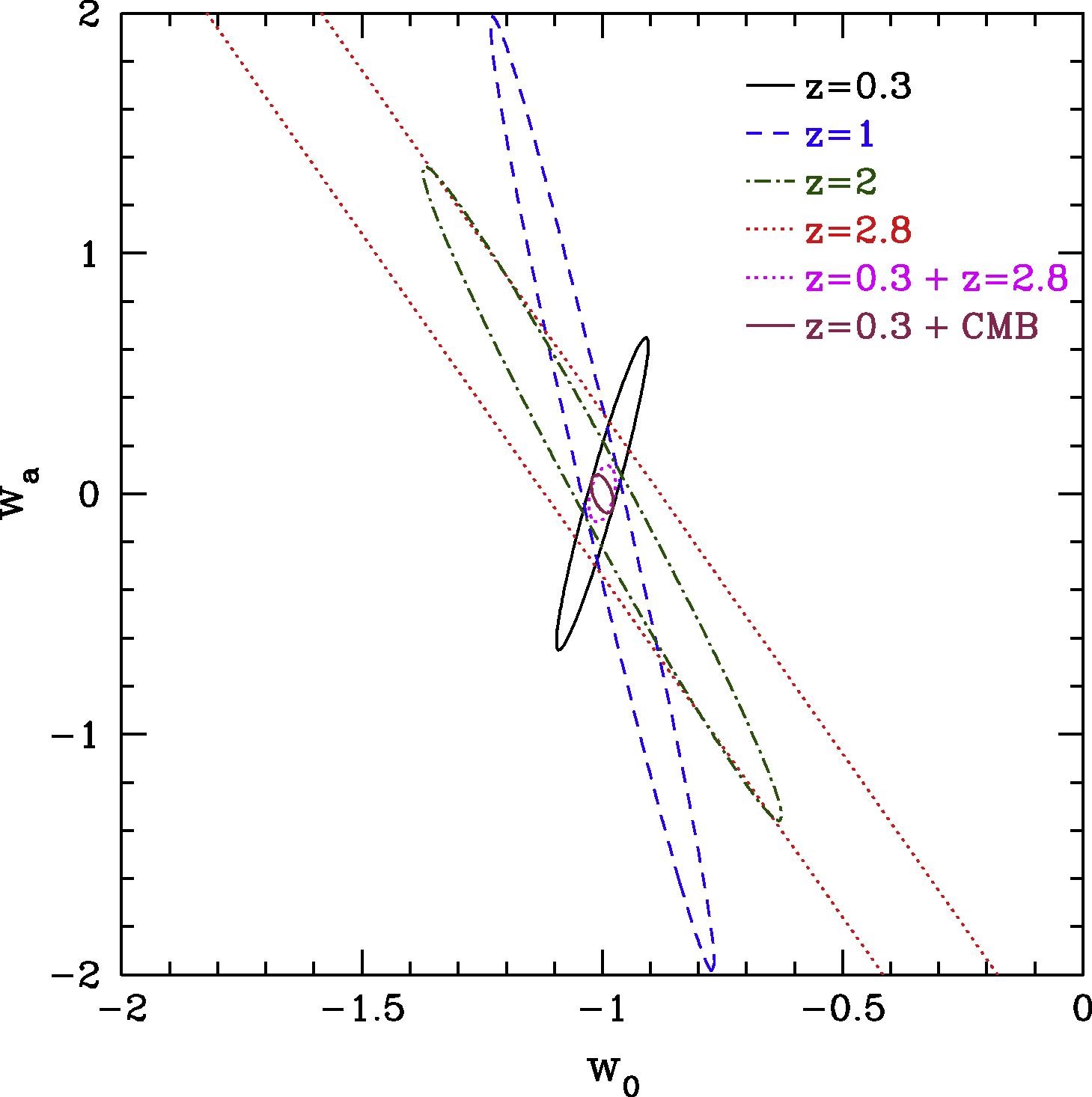}
    \caption{Figure 3 from \cite{kim2015}, which shows the constraints obtained on dark energy equation of state parameters ($w_0$ is the current value; $w_a$ parameterizes evolution in the equation of state) from an experiment that yields five measurements of redshift drift at a given redshift with an accuracy of 1\%. Ellipses correspond to 68\% confidence intervals. Note that low redshift, $z\approx0.3$, is optimal, and has high complementarity with CMB measurements.}
    \label{fig:kim}
\end{figure}

Measurements of cosmic acceleration would not only directly confirm and characterize the Friedmann-Lema{\^{\i}}tre-Robertson-Walker model, but increase the   dark energy probative power (figure of merit) by a factor of 3 beyond
Stage 4 experiments. 
The key new theoretical elements include the redshift optimization analysis indicating measurements at redshift $z\lesssim0.5$ provide the greatest leverage on testing spacetime properties (FLRW) and dark energy (with the further benefit of higher observing signal to noise), and the extraordinary degree of complementarity with CMB measurements at high redshift \citep[see Figure~\ref{fig:kim}, from][]{kim2015}. Meanwhile, early time observations of cosmic acceleration probe the expansion during the decelerating, matter-dominated era. The largest cosmic accelerations are also expected at $z>3$ during this deceleration phase.

The key new experimental elements include controlling systematics by employing differential measurements of wavelengths in an emission line doublet (e.g.\ well characterized [OII]) and interferometric spectroscopy with ultrahigh stability, e.g.\ as enabled by spatial heterodyning and HEP-inspired “crossfading” (optimized weighting). Externally dispersed interferometric (EDI) spectroscopy with crossfading has already demonstrated a factor 1000 gain in stability and reduction of key systematics; the ongoing LLNL LDRD award has further advanced this. Other ideas include laser frequency comb calibration coupled with ensembles of astrophysical calibrators \citep{cosmicaccelerometer},  
and observations at radio wavelengths \citep[e.g.,][]{2020EPJC...80..304L}.  
For example, 
radio observations of the H I 21 cm absorption line observe redshifts of 10 systems in multiple epochs over the course of 13.5 years \citep{2012ApJ...761L..26D}, with reported uncertainty of $O(10^{-8})$, several orders of magnitude
higher than the expected signal. 

Required measurement accuracy for redshift drift is better than a part in $10^{10}$ over a one year baseline (inverse Hubble constant). This corresponds to wavelength shifts equivalent to 1 cm/s/yr, similar to the goal of exoplanet radial velocity experiments 
so there is broad interest in the astrophysics community in enabling this technology\footnote{The LLNL LDRD highlighted the wide applicability of the improvements: in cosmology, exoplanets, fusion research (plasma motions), compact spectroscopy for homeland security applications, and Raman spectroscopy for biomedical imaging.}. One can improve on this in two straightforward ways: longer time baselines (e.g.\ a 5 or 10 year experiment) and many sources or redshift features to reduce the statistical uncertainty below the systematic level. Another necessity is large numbers of photons -- this is helped by the optimum leverage being at low redshifts and from bright (emission line) sources, and the upcoming generation of ELTs. 
Further ideas include a dedicated $\sim$10 meter class telescope or arrays of smaller telescopes. 

In just the first year of the Lawrence Livermore National Laboratory (LLNL) Laboratory Directed Research and Development (LDRD) grant to David Erskine, success of the EDI technique already includes: 
\begin{itemize} 
\item Demonstrated 500--1000$\times$ reduction in wavelength shift systematics 
\item Demonstrated integration with the Keck Planet Finder spectrograph to test wavelength stabilization plus 2$\times$ resolution boost 
\item Demonstrated single delay crossfading with simpler optical design and cancellation of thermal drifts and air convection 
\item Demonstrated stabilization of irregular wavelength-dependent drifts 
\end{itemize} 

Most importantly, all elements seem to be falling into place with the requisite technology now feasible -- and of great interest to other science fields. Experiment and theory have come together to enable one of the most fundamental tests of spacetime and cosmology, finally achievable after 60 years of waiting, in this next decade.

\section{Measuring dark matter sub-structure in the Milky Way}

Measurements of the accelerations of stars give us the most direct window into the mass distributions of galaxies, both the stars and dark matter (the smooth component, as well as dark matter sub-structure).  Traditionally, inferences about the nature of dark matter have been drawn from \emph{estimates} of these accelerations by modeling the positions and velocities of stars, as compared to the distributions of  competing models of dark matter.   For example, self-interacting dark matter (SIDM) is expected to produce a flatter density profile relative to that produced in cold dark matter models (CDM), as dark matter particles scatter elastically with each other at a rate that is quantified by the self-scattering cross-section \citep{Tulin_Yu2018}.  SIDM cosmological simulations also tend to produce more disk-like potentials in Milky-Way type galaxies relative to CDM \citep{Vargya2021}, and a greater diversity in their acceleration and rotation curve profiles \citep{Sameie2020}.  The so-called fuzzy dark matter model which is composed of ultra-light bosons is expected to produce a distinctly different distribution of dark matter on small scales relative to cold dark matter \citep{Hu2000}, i.e., on scales of the de Broglie wavelength (of order a kiloparsec), where it behaves like a wave \citep{Mocz2019,Lancaster2020}.  These small-scale features of competing dark matter models can in principle be constrained from direct acceleration measurements of stars within the Milky Way. 

Kinematic estimates of the acceleration usually rely on assumptions of equilibrium or symmetry that are unlikely to be valid for the Milky Way - and therefore may yield inaccurate descriptions of dark matter in the Milky Way.  Direct measurement of the Galactic acceleration allows us to capture the complexity of the time-dependent Galactic mass distribution (the dark matter and the stars) via extremely precise, time-series observations.  Using the Poisson equation, $\nabla^2 \Phi = - \nabla \cdot \vec a = 4 \pi G \rho $, acceleration measurements can be very straightforwardly be related to the Galactic potential, $\Phi$, and the mass density, $\rho$, without assumption.   The accelerations of stars that live within the gravitational potential of the MW are small (of order $\sim$ 10 cm/s over a decade for stars within $\sim$ kpc of the Sun) but  advances in technology have led to extreme precision spectrographs that can achieve $\sim$ 10 cm/s \citep{Pepe2010,WrightRobertson} and measure the Galactic acceleration directly \citep{Silverwood2019,Chakrabarti2020}.  The current generation of spectrographs like NEID and ESPRESSO have achieved RV precision $\sim$ 10 cm/s, and should enable a measurement of dark matter sub-structure in the Milky Way down to $\sim 10^{9}-10^{10}~M_{\odot}$, as well as the smooth component of the potential, in less than a decade with currently ongoing extreme precision radial velocity surveys.  However, these instruments are on less than 10-m telescopes, and do not access the entire volume of the Milky Way, and are practically limited in the scale of dark matter sub-structure that they can probe.  With the advent of the ELTs - for example spectrographs like G-CLEF on the GMT \citep{GCLEF} and MODHIS on the TMT \citep{Mawet2019}, we can expect to probe the dark matter sub-halo mass function down to $\sim 10^{6}~M_{\odot}$ with measurements across the Galaxy.

Perhaps the most mature precision technique is pulsar timing.  Pulsars are extremely precise Galactic clocks that can be used as accelerometers to measure the Galactic potential. 
Recent analysis of compiled pulsar timing observations of the observed period drift rate of binary pulsars produced the first direct measurement of Galactic accelerations \citep{Chakrabarti2021} using a sample of precisely timed 14 binary pulsars within a $\sim$ kiloparsec of the Sun.  Measurements of these accelerations enabled a determination the mid-plane density (also known as the Oort limit), the local dark matter density, and the shape of the Galactic potential as traced by the pulsars.  

A key signature of fuzzy dark matter that reveals its wave-like nature on scales of the de Broglie wavelength may be detectable by future pulsar timing observations \citep{Porayko2018}.  Expected contributions from future pulsar timing facilities such as the next generation Very Large Array (ng-VLA) and the Deep Synoptic Array (DSA-2000), which will benefit from improvements in sensitivity of about an order of magnitude should enable direct acceleration measurements across the Galaxy and a measurement of dark matter sub-structure, potentially down to the $\sim 10^{6}~M_{\odot}$ scale; these facilities are discussed in the CF3 Facilities White paper "Snowmass 2021 Cosmic Frontier White Paper: Observational Facilities To Study Dark Matter".

Direct acceleration measurements are also now imminently possible by measuring the small shift in the mid-point of the eclipse time (about 0.1 seconds) induced by the Galactic potential of eclipsing binaries observed by \textit{Kepler} about a decade ago \citep{ChakrabartiET}.  These precision measurements are enabled today with \textit{HST}, and in the future with \textit{JWST} and \textit{Roman}.
Planetary astrometric data can also be used to constrain dark matter in the solar system \citep{Pitjev:2013sfa}, study general relativity \citep{DeMarchi:2019lei}, and set new limits on ultralight dark sectors \citep{Tsai:2021irw,Poddar:2020exe}. These studies are extremely crucial for dark matter direct detection studies \citep{Banerjee:2019epw,Banerjee:2019xuy,Tsai:2021lly,Alonso:2022oot} and other areas of fundamental physics studies.


	
\section{High Precision Cosmology Through Cosmological Parallax}

Our motion with respect to the Cosmic Microwave Background (CMB) 
results in positional changes with time for extragalactic sources that depend 
on the transverse co-moving distance. Thus our space motion provides 
a baseline 80 times larger per year than that provided by the Earth's 
annular motion. Since our CMB velocity is both an absolute reference and 
known to better than 1\% the precision, astrometry of quasars can provide a new 
measurement of the Hubble Constant that is independent of the traditional 
methods. Thus, it offers a means to directly address the current tension 
between empirical bootstrapped measurements and the value inferred from 
the consensus standard model (e.g., Freedman 2017). 

The cosmological parallactic distance is related to the transverse co-moving 
distance ($D_M$). Since this secular measure involves a transverse velocity 
it has a different dependence with redshift than does an angular size measure such as 
BAO due to time dilation with redshift. The three standard measures of cosmological 
distance are related through the transverse co-moving distance ($D_M$) but with differing 
dependencies on the cosmological parameters and the Dark Energy equation of state 
(e.g., Weinberg 1971; Hogg 2000; Peebles 2000; Huterer \& Turner 2001). 

The luminosity and angular size distances are well known but the measurement of cosmological 
parallax is less so. Apparent angular positional shifts ($\theta$) due to our motion 
through space can be used to computed the cosmological parallax and the 
transverse co-moving distance: 
$D_P = 1/\theta$. Being purely geometrical the parallax distance minimizes many of the systematics associated with other methods for measuring cosmological geometry. The measurement of cosmological parallax would thus 
provide a new, high precision measure of the cosmological parameters and the Dark Energy 
equation of state that is independent of the methods that have been used 
to date [SN Ia and Baryon Acoustic Oscillations (BAO), e.g., Alam et al. 2020]. 

\subsection{Limitations of the GAIA Astrometric Satellite}

At cosmologically interesting redshifts ($z \sim 1$), the secular parallax for 
the consensus cosmology is approximately $10^{-6}$ arc seconds over ten years 
of our motion. The relatively bright flux limit of the GAIA satellite limits 
precision astrometry to the nearest galactic nuclei and quasars. Furthermore, the presence 
of systematic errors at the few $10^{-6}$ arcsec level limits any parallax measurement via 
ensemble averaging of fainter quasar populations (GAIA Mission Science Performance). 
Various space-based astrometric 
missions have been proposed that may ultimately result in a measurement of cosmological 
parallax using fainter quasars (see Ding \& Croft 2009 for an assessment) but none of these have 
been funded. Similarly, a global array of radio telescopes, such as ngVLA, may ultimately 
achieve the required astrometric precision using a 
sample of compact radio loud quasars (e.g., Paine et al. 2020). The ngVLA was 
endorsed by the Astro 2020 Decadal Survey and received a 2nd priority ranking. 
If funded, it could potentially begin science operations in 2035 and be applied to the 
measurement of cosmological parallax.

\subsection{Precision Cosmology via Astrometry of Gravitationally Lensed Galaxies and Quasars}

An alternative to space-based astrometry and global interferometry is provided by the next 
generation of extremely large telescopes (ELTs) equipped with adaptive optics. These 
telescopes will provide an unprecedented astrometric precision. US participation in two 
of the ELTs currently under development, the Giant Magellan and Thirty Meter Telescopes, 
was ranked the highest priority of Astro 2020 Decadal Review. Most of their funding is 
already in place such that their technical development is underway. Their fields of view will be 
narrow precluding an all sky astrometric survey. However, their narrow-field performance 
still offer distinct advantages for the measurement of cosmological parallax. 
In particular, the gravitational lensing of quasars by foreground galaxies 
magnifies the differential positional shifts between the foreground lens and 
the background source $\sim 5\times$ as our line-of-sight changes with time. Particular 
attention is being paid to the long term astrometric stability of the ELTs. Simulations to 
date imply astrometric precisions of $3 \times 10^{-6}$ arcsec (e.g., Cameron et al. 2009) 
suggesting that these ELTs will provide a measurement of cosmological parallax and the 
transverse co-moving distance for the first time. 

\subsection{Synergies with the Rubin and Roman Surveys and the Path Forward}

The Rubin and Roman surveys are predicted to discover several thousand lensed quasars 
and compact galaxies (Oguri \& Marshall 2010). The angular magnifications in a typical lensed 
systems is about 5x bringing the signal up to a level measurable with the ELTs. Recent simulations 
(Pierce \& McGough, in preparation) have demonstrated that the astrometry of a single system 
with the ELTs will provide a several sigma detection of the cosmological parallax signal. Those 
simulations show that  ELT astrometry of only about 300 systems around the sky would provide 
constraints on the cosmological parameters and the Dark Energy equation of 
state that are comparable to those currently from BAO and SNIa. Thus, the measurement of 
cosmological parallax over the next few decades appears both feasible and inevitable. 

The ELTs promise to provide truly ground breaking capabilities for a number of research 
areas. As a result, the available time on the ELTs will be highly competed. 
We propose two strategies for a precision ELT measurement of the cosmological parallax and 
a corresponding constraint on the transverse co-moving distance:

$\bullet$ {\bf Rely on Public ELT time:} Lensed quasars are relatively bright and simulations 
imply that precision astrometry could be acquired for about 10 systems/night of ELT time. 
Thus precision astrometry for a sample of 300 systems would require approximately 30 nights of 
telescope time for each epoch. The current estimate of the operational cost for the ELTs time is 
\$1.5M/night. This high cost and the extreme competition for public ELT time may limit the measurement of cosmological parallax to a minimal sample measured at two epochs. However, the sample of targets predicted to be discovered with the Rubin and Roman telescopes is sufficiently large to enable higher precision measurements if more ELT time were available.

$\bullet$ {\bf Dedicated ELT Experiment:} The sample of lensed quasars and compact galaxies predicted 
to be found with the Rubin and Roman telescopic surveys could reach several thousand systems. Obtaining precision astrometry for the full sample with public ELT time is likely out of the question.
Thus, to fully leverage the sample of lensed systems for the highest precision measurement of cosmological 
parallax would require a significant, dedicated ELT allocation in both the northern and the southern hemispheres. 
While more expensive, this strategy would also result in constraints on the Dark Energy equation of state 
that are several times higher precision than those currently available from BAO or SNIa.  
A lower precision measurement could be accomplished over a shorter temporal baseline, 
perhaps in only a few years, with a large sample while a corresponding signal could be achieved with a smaller sample over an extended temporal baseline. A dedicated ELT experiment obviously provides the maximum signal. The precision astrometry we are proposing for even the largest samples would require about 10\% of the time available on an ELT.

\section{High precision astrometry with quantum-assisted optical interferometers}

High precision astrometry at the microarcsecond level could open science avenues for imaging black hole accretion disks, improving the local distance ladder, detailing dark matter subhalo influence on microlensing, and dark matter impact visible in Galactic stellar velocity maps. This could be enabled by new ideas cross-cutting the optical interferometry and quantum information science.

Observations using interferometers provide sensitivity to features of images on angular scales much smaller than any single telescope.  Traditional (Michelson stellar) optical interferometers are essentially classical, interfering single photons with themselves \citep{Pedretti2009, Martinod2018, tenBrummelaar2005}, and the single-photon technique is highly developed and approaching technical limits.  Qualitatively new avenues for optical interferometery can be opened up, however, once we consider using multiple-photon states; these generally require a quantum description, especially in conjunction with non-classical quantum technologies such as single-photon sources, entangled pair sources, and quantum memories.  We will focus here on a particular two-photon state technique with application for precision astrometry.

It has been recently proposed that stations in optical interferometers would not require a phase-stable optical link if instead sources of quantum-mechanically entangled pairs could be provided to them, potentially enabling hitherto prohibitively long baselines \citep{Gottesman2012}. If these entangled states could then be interfered locally at each station with an astronomical photon that has impinged on both stations, the single photon counts at the two stations would be correlated in a way that is sensitive to the phase difference in the two paths of the photon, thus reproducing the action of an interferometer.

Several variations of this idea have been proposed. For one of them, which perhaps is a longer term for practical implementation, high intensity wide-band sources of entangled photons and quantum memories would be employed to measure correlations between the stations as explained above \citep{harvard1} . The approach can be generalized from the entanglement of photon pairs to multipartite entanglement in multiple stations and quantum protocols to process information in noisy environment for evaluation of experimental observables. In another approach, which potentially could be implemented in a shorter term, two photons from different sky sources are interfered at two separate stations, requiring only a slow classical information link between them \citep{stankus2020}. The latter scheme can be contrasted with the Hanbury Brown \& Twiss intensity interferometry \citep{hbt} and could allow robust high-precision measurements of the relative astrometry of the two sources. A calculation based on photon statistics suggests that angular precision on the order of $10\mu$as could be achieved in a single night’s observation of two bright stars for a 200m baseline \citep{stankus2020}. We note that this estimate serves only to demonstrate the potential of the technique and is a useful goal for benchmarking in the forthcoming first measurements. Increased sensitivity to fainter objects like galaxies can be achieved for the schemes with bright entangled photon sources and quantum memories \citep{harvard2} employing technologies, which are under development for quantum networks. Though it looks quite futuristic now, the field of quantum information science is going through exponential expansion driven by the industry and, within a decade, may offer capabilities matching the requirements.

Formally, as the baseline is increased to 1,000s km, the projected astrometric resolution could be very small, at the sub-microarcsec level. Of course, the ultimately achievable resolution for this technique remains to be seen as there are important systematics that need to be considered, like the atmospheric turbulence if it is ground based. There is no comprehensive analysis yet of those effects but we note that as a two-photon technique it may have cancellation of uncertainties if the two photons are close enough and propagating through the same atmosphere.

\subsection{Impact on Dark Energy and Dark Matter}

Below we will touch on just a few of the many scientific opportunities afforded by considerable improvements in astrometric precision, which are directly relevant to the dark energy and dark matter studies. 

\textbf{Testing theories of gravity by direct imaging of black hole accretion discs}: The power of intereferometry has recently been demonstrated by the direct imaging of the black hole event horizon in M87 by the Event Horizon Telescope \citep{2019ApJ...875L...2E}. This telescope used the Earth-sized array of telescopes operating in radio bands at 1.1mm to achieve resolution of 25 microarcseconds. Since the telescopes were already spread around earth as much as possible, it is only possible to increase the resolution by using telescopes in space or observing at a smaller wavelength. The quantum-improved techniques advocated here will allow, in principle, for arbitrary baselines, and so by repeating this observation in optical wavelengths it would be possible to increase the resolution by three orders of magnitude (ratio of wavelengths between 1 mm radio and 1 micron optical), bringing about a game changing improvement in resolution. This would open completely new avenues in study of theories of modified gravity that may modify the black hole topology \citep{Moffat2021} and could potentially have large impact on our understanding of dark energy. 

\textbf{Precision parallax and cosmic distance ladder}: 
Significant improvements in astrometric precision would allow for direct parallax measurements of low redshift galaxies hosting Type Ia supernovae. This would enable skipping over several rungs of the local distance ladder, with their prospects for systematic issues, and tie local supernovae more directly to cosmological supernova distance indicators. Direct parallax measurements are systematically
very robust, but are necessarily limited by the achievable astrometric precision. The most sensitive astrometric data
with precision of few dozens microarcsec is provided by the recent Gaia space mission \citep{Katz2019, Lindegren2021}.

\textbf{Mapping microlensing events}: Amongst the candidates for DM are compact star-sized objects, such as black holes, or extended virialized subhalos comprised of yet undiscovered dark matter particles. To probe these DM candidates, a more rigorous and direct method of observing their expected gravitational microlensing effects on stars is needed. The typical photometric measurement of microlensing events both obfuscates details of the lens’s mass and spatial distribution and is less straightforward than an astrometric approach \citep{Erickcek2011, Wyrzykowski2016}. Improvements in astrometric precision would allow for the more direct astrometric approach to mapping microlensing effects and would therefore be beneficial in assessing the viability of certain DM candidates \citep{Grant2021}.

\textbf{Peculiar motions and dark matter}: DM’s effects on the dynamics of our Galaxy are of great interest for understanding its properties and local distribution \citep{Majewski2007, Gardner2021}. The ability to fully measure and reconstruct 3D velocities of a large swath of the stars in the Galaxy would unlock important, thus far inaccessible data that could illuminate many of the unknown characteristics of DM in our galaxy \citep{Steinmetz2020, Simon2018,Katz2019}. Improvements in astrometric measurements are needed to measure the peculiar motion of more distant stars in our Galaxy and subsequently extract their transverse velocity. The improved 3D velocity data afforded by more precise astrometric measurements would pave the way for an inferred measurement of the dark matter halo’s gravitational potential. Moreover, it would allow us to survey historical merging events in the Milky Way halo and open a unique window into DM’s interaction with itself and with ordinary matter \citep{Chu2019}. 

\subsection{Instrument requirements}

An important consideration from the instrumentation viewpoint is that the photons must be close enough in time and in frequency to efficiently interfere; or, formulating it differently, to be indistinguishable within $ \Delta t \cdot \Delta E \sim \hbar $. Converting energy to wavelength, the above is satisfied for $\Delta t \cdot \Delta \lambda = 10~\mathrm{ps} \cdot 0.2~\mathrm{nm}$ at 800~nm wavelength, setting useful target goals for the temporal and spectral resolutions \citep{Nomerotski2020_1}.
Another important parameter for the imaging system is the photon detection efficiency, which needs to be as high as possible, since the two-photon coincidences  have a quadratic dependence on it.

An efficient scheme of spectroscopic binning can be implemented by employing a traditional diffraction grating spectrometer where incoming photons are passed though a slit, dispersed, and then focused onto a linear array of single-photon detectors \citep{Dey2019, Vogt1994, Zhang2020}. However, improvement of timing resolution appears to be the most straightforward way to achieve the targeted performance. Fast technologies, such as superconducting nanowire single photon detectors (SNSPD) and single photon avalanche devices (SPAD), can be considered for this application. The superconducting nanowire detectors have excellent photon detection efficiency, in excess of 90\% \citep{Zhu2020, Divochiy2008}, with demonstrated 3~ps timing resolution for single devices~\citep{Korzh2020}.
The SPAD sensors are based on silicon diodes with engineered junction breakdown producing fast pulses of big enough amplitude for single photons. These devices also have excellent timing resolution, which can be as good as 10~ps for single-channel devices, and most importantly, good potential for scalability with multi-channel imagers already reported \citep{Gasparini2017, Morimoto2020}. Benchmarking of these promising technologies for a spectrograph with required spectral and timing resolutions is currently in progress \citep{nomerotski2021}.

\begin{center}
Table 1: Precision Frontiers

\begin{tabular}{ |p{3cm}||p{4.5cm}|p{3.5cm}|p{4cm}|  }
\hline
Precision Frontier & Science & Key Technologies & Technology Status \\
 \hline
 Extreme Precision & Redshift Drift (Dark Energy) & EDI Spectroscopy & Deployed Prototype \\ 
  Radial Velocity & Dark Matter Substructure & Actively Stabilized Spectrographs & Designed, untested \\
  & & Dedicated large-aperture facility & Single aperture technology mature; prototype exist for telescope arrays \\
  \hline
 Astrometry & Cosmological Parallax |& ELT-class telescopes & N/A \\
 & Distance Ladder & Quantum-assisted  & In development \\
 & Dark Matter Substructure &optical interferometers & \\
 \hline
\end{tabular}
\end{center}

\section{Technology Status and the Path Forward}

\begin{figure}[h]        
\begin{center}
\includegraphics[scale=0.6]{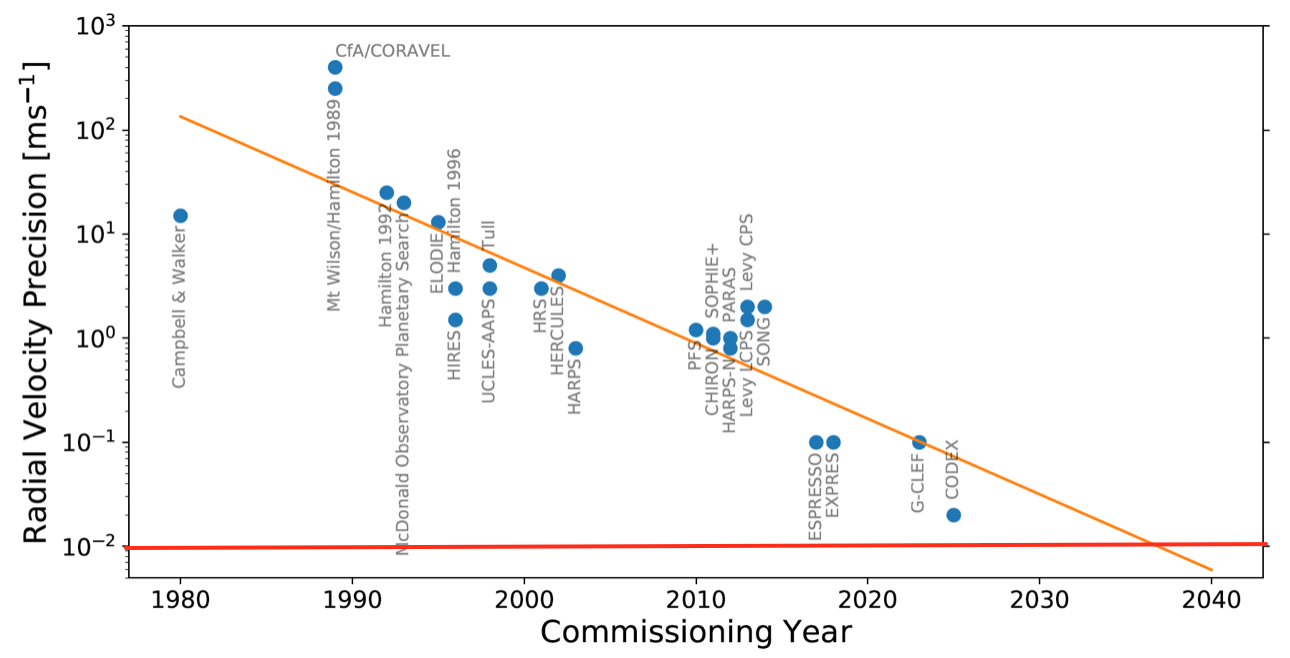}
\caption{Improvements in RV precision for various upcoming instruments (adapted from Silverwood \& Easther 2019). Cosmological redshift drift requires $\sim 1$ cm s$^{-1}$ precision (red line) with stability of years to decades.}
\label{f:RVprecision}
\end{center}
\end{figure}

A unifying theme with the above science cases is that they are all now technologically feasible. Table 1 collates the key science cases associated with high-precision spectroscopic velocity measurements and high-precision astrometric measurements, and summarizes the status of key technologies associated with each.

For radial velocities, the next generation of instruments must achieve $<1$ cm s$^{-1}$ velocity precision with stability on timescales of years to decades, and must be deployed at facilities with 10m to ELT-class apertures (depending on the redshift probed).  Of currently planned spectrographs, G-CLEF for the GMT and MODHIS for the TMT represent the state-of-the-art and are designed to yield 10 cm s$^{-1}$ precision \citep{GCLEF,Mawet2019}.

There are two promising techniques that have been suggested for achieving the required precision and stability. The earlier generation of instruments have already demonstrated RV precision; the remaining technical challenge is now demonstrating stability over the decade timescales that are necessary to measure dark matter and dark energy.  One technique for achieving the requisite stability is the crossfading method for externally dispersed interferometric spectroscopy  \citep{2020SPIE11451E..2DE,2021JATIS...7b5006E}. This approach has been demonstrated on-sky to yield a $10^3$ gain in stability \citep{2021JATIS...7b5006E,2020SPIE11451E..2DE}, and if used with doublet lines for redshift drift yields a differential measurement that mitigates experimental systematics.

The second technique is the use of active stabilization of spectrographs for extreme precision radial velocities. For a standard high-resolution spectrograph any changes in the optomechanics due to environment or other factors induce a corresponding drift in where light is dispersed onto the detector. Even if the changes are at the sub-pixel level, such shifts can induce systematic biases in the recovered velocities due to sub-pixel detector physics. The most recent generation of high-precisions spectrographs such as NEID \citep{NEID_optical} include thermal stabilization techniques to mitigate these factors, however this mitigation is still passive. An alternative that has been proposed \citep{cosmicaccelerometer} is to incorporate LIGO technology, using a laser cavity to continuously measure the optical dimensions of the spectrograph and feed these into an active control loop with thermal heaters directly coupled to the optical bench of the spectrograph. The components of this technology are demonstrated, but no prototype of an actively stabilized spectrograph exists at this time.
The above technologies are also applicable to radial velocity measurements for galactic sub-structure; however for this science the RV precision requirements are less severe. The G-CLEF and MODHIS instruments on the GMT and TMT respectively, which are designed for a velocity precision of $10$ cm s$^{-1}$, enable this science, at least to constrain the sub-halo mass function on scales greater than $10^{6}~M_{\odot}$.  To constrain the low-mass end of the sub-halo mass function would require $\sim$ cm/s RV precision. 

These spectrographs must then be deployed on large-aperture telescopes. Deployment of the EDI technology for low-redshift doublet measurements requires 10m-class telescopes \citep{2020SPIE11451E..2DE}, while the active spectrograph design coupled to 20m-class telescopes has been suggested for high-redshift Lyman-forest redshift drift measurements.\footnote{See \citet{Liske2008} for more information on Lyman-forest redshift drift measurements, which were first suggested by \citet{loeb1998}.} The dark matter sub-structure science also requires $>$ 10m-class telescopes to enable sufficient SNR to allow for acceleration measurements across the Galaxy. While one option is the use of current and upcoming telescopes, the ELTs in particularly will be highly oversubscribed when they come online, with a number of competing science priorities. A dedicated facility for high-precision astrophysics has multiple advantages. These include the ability to optimize the design to robustly control systematics (including those which may arise from switches between various instruments at a general use facility), a guaranteed large amount of observing time for these investigations, and the ability optimize the experimental setup for calibrations and optimal precision. As an alternative to a traditional telescope design, \cite{eikenberry2019b} have suggested the construction of an array of small, fiber-coupled telescopes that feed into a single spectrograph via photonic lanterns. The authors argue that such an approach can be a factor of ten lower cost than a traditional design, and is compatible with EDI technology. At present, this team is constructing a small demonstrator array for deployment at Mt.~Laguna Observatory by early 2023, which will validate field deployment of photonic lantern technology and the array design for spectroscopy.

For astrometry, one of the key sciences (cosmological parallaxes) will be feasible with planned instrumentation on upcoming ELTs with no additional facilities or technical development. However, a dedicated ELT experiment may be required to maximize the cosmological parallax signal over the shortest temporal baseline. Another promising venue for astrometry relies upon development of quantum-assisted optical interferometers to enable $\mu$as-scale astrometric precision, potentially even for ground-based installations. As noted in section 5, this technology is at an R \& D state at present and will require dedicated resources for continued development and deployment of technology demonstrators before it can be deployed to address the core science. 

Given this frontier for high-precision cosmological and Galactic radial velocity, and astrometric, measurements the path towards enabling deployment of these technologies requires a combination of instrumentation R \& D -- with the maturity level of the technologies varying from the design phase to demonstrated prototypes -- and securing access to large aperture facilities to obtain sufficient sensitivity for the proposed measurements. 

\bibliographystyle{aasjournal}
\bibliography{bibl}

\begin{thebibliography}{}
\expandafter\ifx\csname natexlab\endcsname\relax\def\natexlab#1{#1}\fi
\providecommand{\url}[1]{\href{#1}{#1}}
\providecommand{\dodoi}[1]{doi:~\href{http://doi.org/#1}{\nolinkurl{#1}}}
\providecommand{\doeprint}[1]{\href{http://ascl.net/#1}{\nolinkurl{http://ascl.net/#1}}}
\providecommand{\doarXiv}[1]{\href{https://arxiv.org/abs/#1}{\nolinkurl{https://arxiv.org/abs/#1}}}

\bibitem[{Gra(2021)}]{Grant2021}
 2021, Physics Today, 2021, 1221a, \dodoi{10.1063/pt.6.1.20211221a}

\bibitem[{{Alam} {et~al.}(2021){Alam}, {Aubert}, {Avila}, {Balland},
  {Bautista}, {Bershady}, {Bizyaev}, {Blanton}, {Bolton}, {Bovy}, {Brinkmann},
  {Brownstein}, {Burtin}, {Chabanier}, {Chapman}, {Choi}, {Chuang}, {Comparat},
  {Cousinou}, {Cuceu}, {Dawson}, {de la Torre}, {de Mattia}, {Agathe}, {des
  Bourboux}, {Escoffier}, {Etourneau}, {Farr}, {Font-Ribera}, {Frinchaboy},
  {Fromenteau}, {Gil-Mar{\'\i}n}, {Le Goff}, {Gonzalez-Morales},
  {Gonzalez-Perez}, {Grabowski}, {Guy}, {Hawken}, {Hou}, {Kong}, {Parker},
  {Klaene}, {Kneib}, {Lin}, {Long}, {Lyke}, {de la Macorra}, {Martini},
  {Masters}, {Mohammad}, {Moon}, {Mueller}, {Mu{\~n}oz-Guti{\'e}rrez}, {Myers},
  {Nadathur}, {Neveux}, {Newman}, {Noterdaeme}, {Oravetz}, {Oravetz},
  {Palanque-Delabrouille}, {Pan}, {Paviot}, {Percival}, {P{\'e}rez-R{\`a}fols},
  {Petitjean}, {Pieri}, {Prakash}, {Raichoor}, {Ravoux}, {Rezaie}, {Rich},
  {Ross}, {Rossi}, {Ruggeri}, {Ruhlmann-Kleider}, {S{\'a}nchez}, {S{\'a}nchez},
  {S{\'a}nchez-Gallego}, {Sayres}, {Schneider}, {Seo}, {Shafieloo}, {Slosar},
  {Smith}, {Stermer}, {Tamone}, {Tinker}, {Tojeiro}, {Vargas-Maga{\~n}a},
  {Variu}, {Wang}, {Weaver}, {Weijmans}, {Y{\`e}che}, {Zarrouk}, {Zhao},
  {Zhao}, \& {Zheng}}]{alam2021}
{Alam}, S., {Aubert}, M., {Avila}, S., {et~al.} 2021, \prd, 103, 083533,
  \dodoi{10.1103/PhysRevD.103.083533}

\bibitem[{Alonso {et~al.}(2022)}]{Alonso:2022oot}
Alonso, I., {et~al.} 2022, in .
\newblock \doarXiv{2201.07789}

\bibitem[{Banerjee {et~al.}(2020{\natexlab{a}})Banerjee, Budker, Eby, Flambaum,
  Kim, Matsedonskyi, \& Perez}]{Banerjee:2019xuy}
Banerjee, A., Budker, D., Eby, J., {et~al.} 2020{\natexlab{a}}, JHEP, 09, 004,
  \dodoi{10.1007/JHEP09(2020)004}

\bibitem[{Banerjee {et~al.}(2020{\natexlab{b}})Banerjee, Budker, Eby, Kim, \&
  Perez}]{Banerjee:2019epw}
Banerjee, A., Budker, D., Eby, J., Kim, H., \& Perez, G. 2020{\natexlab{b}},
  Commun. Phys., 3, 1, \dodoi{10.1038/s42005-019-0260-3}

\bibitem[{Brown \& Twiss(1956)}]{hbt}
Brown, R.~H., \& Twiss, R.~Q. 1956, Nature, 178, 1046,
  \dodoi{10.1038/1781046a0}

\bibitem[{{Chakrabarti} {et~al.}(2021{\natexlab{a}}){Chakrabarti}, {Chang},
  {Lam}, {Vigeland}, \& {Quillen}}]{Chakrabarti2021}
{Chakrabarti}, S., {Chang}, P., {Lam}, M.~T., {Vigeland}, S.~J., \& {Quillen},
  A.~C. 2021{\natexlab{a}}, \apjl, 907, L26, \dodoi{10.3847/2041-8213/abd635}

\bibitem[{{Chakrabarti} {et~al.}(2021{\natexlab{b}}){Chakrabarti}, {Stevens},
  {Wright}, {Rafikov}, {Chang}, {Beatty}, \& {Huber}}]{ChakrabartiET}
{Chakrabarti}, S., {Stevens}, D.~J., {Wright}, J., {et~al.} 2021{\natexlab{b}},
  arXiv e-prints, arXiv:2112.08231.
\newblock \doarXiv{2112.08231}

\bibitem[{{Chakrabarti} {et~al.}(2020){Chakrabarti}, {Wright}, {Chang},
  {Quillen}, {Craig}, {Territo}, {D'Onghia}, {Johnston}, {De Rosa}, {Huber},
  {Rhode}, \& {Nielsen}}]{Chakrabarti2020}
{Chakrabarti}, S., {Wright}, J., {Chang}, P., {et~al.} 2020, arXiv e-prints,
  arXiv:2007.15097.
\newblock \doarXiv{2007.15097}

\bibitem[{Chu {et~al.}(2019)Chu, Garcia-Cely, \& Murayama}]{Chu2019}
Chu, X., Garcia-Cely, C., \& Murayama, H. 2019, Phys. Rev. Lett., 122, 071103,
  \dodoi{10.1103/PhysRevLett.122.071103}

\bibitem[{{Darling}(2012)}]{2012ApJ...761L..26D}
{Darling}, J. 2012, \apjl, 761, L26, \dodoi{10.1088/2041-8205/761/2/L26}

\bibitem[{De~Marchi \& Cascioli(2020)}]{DeMarchi:2019lei}
De~Marchi, F., \& Cascioli, G. 2020, Class. Quant. Grav., 37, 095007,
  \dodoi{10.1088/1361-6382/ab6ae0}

\bibitem[{Dey {et~al.}(2019)Dey, Schlegel, Lang, Blum, Burleigh, Fan, Findlay,
  Finkbeiner, Herrera, Juneau, Landriau, Levi, McGreer, Meisner,
  {et~al.}}]{Dey2019}
Dey, A., Schlegel, D.~J., Lang, D., {et~al.} 2019, The Astronomical Journal,
  157, 168, \dodoi{10.3847/1538-3881/ab089d}

\bibitem[{Divochiy {et~al.}(2008)Divochiy, Marsili, Bitauld, Gaggero, Leoni,
  Mattioli, Korneev, Seleznev, Kaurova, Minaeva, Gol{\textquotesingle}tsman,
  Lagoudakis, Benkhaoul, L{\'{e}}vy, \& Fiore}]{Divochiy2008}
Divochiy, A., Marsili, F., Bitauld, D., {et~al.} 2008, Nature Photonics, 2,
  302, \dodoi{10.1038/nphoton.2008.51}

\bibitem[{{Eikenberry} {et~al.}(2019{\natexlab{a}}){Eikenberry}, {Gonzalez},
  {Darling}, {Liske}, {Slepian}, {Mueller}, {Conklin}, {Fulda}, {Mendes de
  Oliveira}, {Bentz}, {Jeram}, {Dong}, {Townsend}, {Izuti Nakazono}, {Quimby},
  \& {Welsh}}]{cosmicaccelerometer}
{Eikenberry}, S., {Gonzalez}, A., {Darling}, J., {et~al.} 2019{\natexlab{a}},
  in Bulletin of the American Astronomical Society, Vol.~51, 137.
\newblock \doarXiv{1907.08271}

\bibitem[{{Eikenberry} {et~al.}(2019{\natexlab{b}}){Eikenberry}, {Bentz},
  {Gonzalez}, {Harrington}, {Jeram}, {Law}, {Maccarone}, {Quimby}, \&
  {Townsend}}]{eikenberry2019b}
{Eikenberry}, S., {Bentz}, M., {Gonzalez}, A., {et~al.} 2019{\natexlab{b}}, in
  Bulletin of the American Astronomical Society, Vol.~51, 124.
\newblock \doarXiv{1907.08273}

\bibitem[{Erickcek \& Law(2011)}]{Erickcek2011}
Erickcek, A.~L., \& Law, N.~M. 2011, The Astrophysical Journal, 729, 49,
  \dodoi{10.1088/0004-637x/729/1/49}

\bibitem[{{Erskine}(2020)}]{2020SPIE11451E..2DE}
{Erskine}, D.~J. 2020, in Society of Photo-Optical Instrumentation Engineers
  (SPIE) Conference Series, Vol. 11451, Society of Photo-Optical
  Instrumentation Engineers (SPIE) Conference Series, 114512D,
  \dodoi{10.1117/12.2559219}

\bibitem[{{Erskine}(2021)}]{2021JATIS...7b5006E}
{Erskine}, D.~J. 2021, Journal of Astronomical Telescopes, Instruments, and
  Systems, 7, 025006, \dodoi{10.1117/1.JATIS.7.2.025006}

\bibitem[{{Event Horizon Telescope Collaboration} {et~al.}(2019){Event Horizon
  Telescope Collaboration}, {Akiyama}, {Alberdi}, {Alef}, {Asada}, {Azulay},
  {Baczko}, {Ball}, {Balokovi{\'c}}, {Barrett}, {Bintley},
  {et~al.}}]{2019ApJ...875L...2E}
{Event Horizon Telescope Collaboration}, {Akiyama}, K., {Alberdi}, A., {et~al.}
  2019, \apjl, 875, L2, \dodoi{10.3847/2041-8213/ab0c96}

\bibitem[{Gardner {et~al.}(2021)Gardner, McDermott, \& Yanny}]{Gardner2021}
Gardner, S., McDermott, S.~D., \& Yanny, B. 2021, Progress in Particle and
  Nuclear Physics, 121, 103904, \dodoi{10.1016/j.ppnp.2021.103904}

\bibitem[{Gasparini {et~al.}(2017)Gasparini, Bessire, Untern\"{a}hrer,
  Stefanov, Boiko, Perenzoni, \& Stoppa}]{Gasparini2017}
Gasparini, L., Bessire, B., Untern\"{a}hrer, M., {et~al.} 2017, in Quantum
  Sensing and Nano Electronics and Photonics {XIV}, ed. M.~Razeghi ({SPIE}),
  \dodoi{10.1117/12.2253598}

\bibitem[{Gottesman {et~al.}(2012)Gottesman, Jennewein, \&
  Croke}]{Gottesman2012}
Gottesman, D., Jennewein, T., \& Croke, S. 2012, Phys. Rev. Lett., 109, 070503,
  \dodoi{10.1103/PhysRevLett.109.070503}

\bibitem[{{Hu} {et~al.}(2000){Hu}, {Barkana}, \& {Gruzinov}}]{Hu2000}
{Hu}, W., {Barkana}, R., \& {Gruzinov}, A. 2000, \prl, 85, 1158,
  \dodoi{10.1103/PhysRevLett.85.1158}

\bibitem[{Katz {et~al.}(2019)Katz, Sartoretti, Cropper, Panuzzo, Seabroke,
  Viala, Benson, Blomme, Jasniewicz, Jean-Antoine, Huckle, Smith, Baker, Crifo,
  Damerdji, David, Dolding, Fr{\'{e}}mat, Gosset, Guerrier, Guy, Haigron,
  Jan{\ss}en, Marchal, Plum, Soubiran, Th{\'{e}}venin, Ajaj, Prieto, Babusiaux,
  Boudreault, Chemin, Luche, Fabre, Gueguen, Hambly, Lasne, Meynadier, Pailler,
  Panem, Royer, Tauran, Zurbach, Zwitter, Arenou, Bossini, Gerssen,
  G{\'{o}}mez, Lemaitre, Leclerc, Morel, Munari, Turon, Vallenari, \&
  {\v{Z}}erjal}]{Katz2019}
Katz, D., Sartoretti, P., Cropper, M., {et~al.} 2019, Astronomy {\&}
  Astrophysics, 622, A205, \dodoi{10.1051/0004-6361/201833273}

\bibitem[{Khabiboulline {et~al.}(2019{\natexlab{a}})Khabiboulline, Borregaard,
  De~Greve, \& Lukin}]{harvard1}
Khabiboulline, E.~T., Borregaard, J., De~Greve, K., \& Lukin, M.~D.
  2019{\natexlab{a}}, Phys. Rev. A, 100, 022316,
  \dodoi{10.1103/PhysRevA.100.022316}

\bibitem[{Khabiboulline {et~al.}(2019{\natexlab{b}})Khabiboulline, Borregaard,
  De~Greve, \& Lukin}]{harvard2}
---. 2019{\natexlab{b}}, Phys. Rev. Lett., 123, 070504,
  \dodoi{10.1103/PhysRevLett.123.070504}

\bibitem[{{Kim} {et~al.}(2015){Kim}, {Linder}, {Edelstein}, \&
  {Erskine}}]{kim2015}
{Kim}, A.~G., {Linder}, E.~V., {Edelstein}, J., \& {Erskine}, D. 2015,
  Astroparticle Physics, 62, 195, \dodoi{10.1016/j.astropartphys.2014.09.004}

\bibitem[{Korzh {et~al.}(2020)Korzh, Zhao, Allmaras, Frasca, Autry, Bersin,
  Beyer, Briggs, Bumble, Colangelo, Crouch, Dane, Gerrits, Lita, Marsili,
  Moody, Pe{\~{n}}a, Ramirez, Rezac, Sinclair, Stevens, Velasco, Verma,
  Wollman, Xie, Zhu, Hale, Spiropulu, Silverman, Mirin, Nam, Kozorezov, Shaw,
  \& Berggren}]{Korzh2020}
Korzh, B., Zhao, Q.-Y., Allmaras, J.~P., {et~al.} 2020, Nature Photonics, 14,
  250, \dodoi{10.1038/s41566-020-0589-x}

\bibitem[{Kumar~Poddar {et~al.}(2020)Kumar~Poddar, Mohanty, \&
  Jana}]{Poddar:2020exe}
Kumar~Poddar, T., Mohanty, S., \& Jana, S. 2020.
\newblock \doarXiv{2002.02935}

\bibitem[{{Lancaster} {et~al.}(2020){Lancaster}, {Giovanetti}, {Mocz}, {Kahn},
  {Lisanti}, \& {Spergel}}]{Lancaster2020}
{Lancaster}, L., {Giovanetti}, C., {Mocz}, P., {et~al.} 2020, \jcap, 2020, 001,
  \dodoi{10.1088/1475-7516/2020/01/001}

\bibitem[{Lindegren {et~al.}(2021)Lindegren, Klioner, Hern{\'{a}}ndez, Bombrun,
  Ramos-Lerate, Steidelm\"{u}ller, Bastian, Biermann, de~Torres, Gerlach,
  Geyer, Hilger, Hobbs, Lammers, McMillan, Stephenson, Casta{\~{n}}eda,
  Davidson, Fabricius, Gracia-Abril, Portell, Rowell, Teyssier, Torra,
  Bartolom{\'{e}}, Clotet, Garralda, Gonz{\'{a}}lez-Vidal, Torra, Abbas,
  Altmann, Varela, Balaguer-N{\'{u}}{\~{n}}ez, Balog, Barache, Becciani,
  Bernet, Bertone, Bianchi, Bouquillon, Brown, Bucciarelli, Busonero,
  Butkevich, Buzzi, Cancelliere, Carlucci, Charlot, Cioni, Crosta, Crowley, del
  Peloso, del Pozo, Drimmel, Esquej, Fienga, Fraile, Gai, Garcia-Reinaldos,
  Guerra, Hambly, Hauser, Jan{\ss}en, Jordan, Kostrzewa-Rutkowska, Lattanzi,
  Liao, Licata, Lister, L\"{o}ffler, Marchant, Masip, Mignard, Mints, Molina,
  Mora, Morbidelli, Murphy, Pagani, Panuzzo, Esteller, Poggio, Fiorentin, Riva,
  Sell{\'{e}}s, Gimenez, Sarasso, Sciacca, Siddiqui, Smart, Souami, Spagna,
  Steele, Taris, Utrilla, van Reeven, \& Vecchiato}]{Lindegren2021}
Lindegren, L., Klioner, S.~A., Hern{\'{a}}ndez, J., {et~al.} 2021, Astronomy
  {\&} Astrophysics, 649, A2, \dodoi{10.1051/0004-6361/202039709}

\bibitem[{{Liske} {et~al.}(2008){Liske}, {Grazian}, {Vanzella}, {Dessauges},
  {Viel}, {Pasquini}, {Haehnelt}, {Cristiani}, {Pepe}, {Avila}, {Bonifacio},
  {Bouchy}, {Dekker}, {Delabre}, {D'Odorico}, {D'Odorico}, {Levshakov},
  {Lovis}, {Mayor}, {Molaro}, {Moscardini}, {Murphy}, {Queloz}, {Shaver},
  {Udry}, {Wiklind}, \& {Zucker}}]{Liske2008}
{Liske}, J., {Grazian}, A., {Vanzella}, E., {et~al.} 2008, \mnras, 386, 1192,
  \dodoi{10.1111/j.1365-2966.2008.13090.x}

\bibitem[{{Liu} {et~al.}(2020){Liu}, {Zhang}, \& {Zhang}}]{2020EPJC...80..304L}
{Liu}, Y., {Zhang}, J.-F., \& {Zhang}, X. 2020, European Physical Journal C,
  80, 304, \dodoi{10.1140/epjc/s10052-020-7863-4}

\bibitem[{{Loeb}(1998)}]{loeb1998}
{Loeb}, A. 1998, \apjl, 499, L111, \dodoi{10.1086/311375}

\bibitem[{Majewski(2007)}]{Majewski2007}
Majewski, S.~R. 2007, Proceedings of the International Astronomical Union, 3,
  450, \dodoi{10.1017/s1743921308019790}

\bibitem[{Martinod {et~al.}(2018)Martinod, Mourard, B{\'{e}}rio, Perraut,
  Meilland, Bailet, Bresson, ten Brummelaar, Clausse, Dejonghe, Ireland,
  Millour, Monnier, Sturmann, Sturmann, \& Tallon}]{Martinod2018}
Martinod, M.~A., Mourard, D., B{\'{e}}rio, P., {et~al.} 2018, Astronomy {\&}
  Astrophysics, 618, A153, \dodoi{10.1051/0004-6361/201731386}

\bibitem[{{Mawet} {et~al.}(2019){Mawet}, {Fitzgerald}, {Konopacky}, {Beichman},
  {Jovanovic}, {Dekany}, {Hover}, {Chisholm}, {Ciardi}, {Artigau}, {Banyal},
  {Beatty}, {Benneke}, {Blake}, {Burgasser}, {Canalizo}, {Chen}, {Do},
  {Doppmann}, {Doyon}, {Dressing}, {Fang}, {Greene}, {Hillenbrand}, {Howard},
  {Kane}, {Kataria}, {Kempton}, {Knutson}, {Kotani}, {Lafreni{\`e}re}, {Liu},
  {Nishiyama}, {Pandey}, {Plavchan}, {Prato}, {Rajaguru}, {Robertson}, {Salyk},
  {Sato}, {Schlawin}, {Sengupta}, {Sivarani}, {Skidmore}, {Tamura}, {Terada},
  {Vasisht}, {Wang}, \& {Zhang}}]{Mawet2019}
{Mawet}, D., {Fitzgerald}, M., {Konopacky}, Q., {et~al.} 2019, in Bulletin of
  the American Astronomical Society, Vol.~51, 134.
\newblock \doarXiv{1908.03623}

\bibitem[{{McVittie}(1962)}]{mcvittie1962}
{McVittie}, G.~C. 1962, \apj, 136, 334

\bibitem[{{Mocz} {et~al.}(2019){Mocz}, {Fialkov}, {Vogelsberger}, {Becerra},
  {Amin}, {Bose}, {Boylan-Kolchin}, {Chavanis}, {Hernquist}, {Lancaster},
  {Marinacci}, {Robles}, \& {Zavala}}]{Mocz2019}
{Mocz}, P., {Fialkov}, A., {Vogelsberger}, M., {et~al.} 2019, \prl, 123,
  141301, \dodoi{10.1103/PhysRevLett.123.141301}

\bibitem[{Moffat(2021)}]{Moffat2021}
Moffat, J. 2021, Journal of Cosmology and Astroparticle Physics, 2021, 017,
  \dodoi{10.1088/1475-7516/2021/02/017}

\bibitem[{Morimoto {et~al.}(2020)Morimoto, Ardelean, Wu, Ulku, Antolovic,
  Bruschini, \& Charbon}]{Morimoto2020}
Morimoto, K., Ardelean, A., Wu, M.-L., {et~al.} 2020, Optica, 7, 346,
  \dodoi{10.1364/optica.386574}

\bibitem[{Nomerotski {et~al.}(2021)Nomerotski, Schiff, Stankus, Keach,
  Parsells, Saira, Slo{\v{z}}ar, \& Vintskevich}]{nomerotski2021}
Nomerotski, A., Schiff, J., Stankus, P., {et~al.} 2021, arXiv preprint
  arXiv:2107.09229

\bibitem[{Nomerotski {et~al.}(2020)Nomerotski, Stankus, Solar, Vintskevich,
  Andrewski, Carini, Dolzhenko, England, Figueroa, Gera, Haupt, Herrmann,
  Katramatos, Keach, Parsells, Saira, Schiff, Svihra, Tsang, \&
  Zhang}]{Nomerotski2020_1}
Nomerotski, A., Stankus, P., Solar, A., {et~al.} 2020, in Optical and Infrared
  Interferometry and Imaging {VII}, ed. A.~M{\'{e}}rand, S.~Sallum, \& P.~G.
  Tuthill ({SPIE}), \dodoi{10.1117/12.2560272}

\bibitem[{Pedretti {et~al.}(2009)Pedretti, Monnier, ten Brummelaar, \&
  Thureau}]{Pedretti2009}
Pedretti, E., Monnier, J.~D., ten Brummelaar, T., \& Thureau, N.~D. 2009, New
  Astronomy Reviews, 53, 353, \dodoi{10.1016/j.newar.2010.07.008}

\bibitem[{{Pepe} {et~al.}(2021){Pepe}, {Cristiani}, {Rebolo}, {Santos},
  {Dekker}, {Cabral}, {Di Marcantonio}, {Figueira}, {Lo Curto}, {Lovis},
  {Mayor}, {M{\'e}gevand}, {Molaro}, {Riva}, {Zapatero Osorio}, {Amate},
  {Manescau}, {Pasquini}, {Zerbi}, {Adibekyan}, {Abreu}, {Affolter}, {Alibert},
  {Aliverti}, {Allart}, {Allende Prieto}, {{\'A}lvarez}, {Alves}, {Avila},
  {Baldini}, {Bandy}, {Barros}, {Benz}, {Bianco}, {Borsa}, {Bourrier},
  {Bouchy}, {Broeg}, {Calderone}, {Cirami}, {Coelho}, {Conconi}, {Coretti},
  {Cumani}, {Cupani}, {D'Odorico}, {Damasso}, {Deiries}, {Delabre},
  {Demangeon}, {Dumusque}, {Ehrenreich}, {Faria}, {Fragoso}, {Genolet},
  {Genoni}, {G{\'e}nova Santos}, {Gonz{\'a}lez Hern{\'a}ndez}, {Hughes},
  {Iwert}, {Kerber}, {Knudstrup}, {Landoni}, {Lavie}, {Lillo-Box}, {Lizon},
  {Maire}, {Martins}, {Mehner}, {Micela}, {Modigliani}, {Monteiro}, {Monteiro},
  {Moschetti}, {Murphy}, {Nunes}, {Oggioni}, {Oliveira}, {Oshagh}, {Pall{\'e}},
  {Pariani}, {Poretti}, {Rasilla}, {Rebord{\~a}o}, {Redaelli}, {Santana
  Tschudi}, {Santin}, {Santos}, {S{\'e}gransan}, {Schmidt}, {Segovia},
  {Sosnowska}, {Sozzetti}, {Sousa}, {Span{\`o}}, {Su{\'a}rez Mascare{\~n}o},
  {Tabernero}, {Tenegi}, {Udry}, \& {Zanutta}}]{Pepe2021}
{Pepe}, F., {Cristiani}, S., {Rebolo}, R., {et~al.} 2021, \aap, 645, A96,
  \dodoi{10.1051/0004-6361/202038306}

\bibitem[{{Pepe} {et~al.}(2010){Pepe}, {Cristiani}, {Rebolo Lopez}, {Santos},
  {Amorim}, {Avila}, {Benz}, {Bonifacio}, {Cabral}, {Carvas}, {Cirami},
  {Coelho}, {Comari}, {Coretti}, {De Caprio}, {Dekker}, {Delabre}, {Di
  Marcantonio}, {D'Odorico}, {Fleury}, {Garc{\'\i}a}, {Herreros Linares},
  {Hughes}, {Iwert}, {Lima}, {Lizon}, {Lo Curto}, {Lovis}, {Manescau},
  {Martins}, {M{\'e}gevand}, {Moitinho}, {Molaro}, {Monteiro}, {Monteiro},
  {Pasquini}, {Mordasini}, {Queloz}, {Rasilla}, {Rebord{\~a}o}, {Santana
  Tschudi}, {Santin}, {Sosnowska}, {Span{\`o}}, {Tenegi}, {Udry}, {Vanzella},
  {Viel}, {Zapatero Osorio}, \& {Zerbi}}]{Pepe2010}
{Pepe}, F.~A., {Cristiani}, S., {Rebolo Lopez}, R., {et~al.} 2010, in Society
  of Photo-Optical Instrumentation Engineers (SPIE) Conference Series, Vol.
  7735, \procspie, 77350F, \dodoi{10.1117/12.857122}

\bibitem[{Pitjev \& Pitjeva(2013)}]{Pitjev:2013sfa}
Pitjev, N.~P., \& Pitjeva, E.~V. 2013, Astron. Lett., 39, 141,
  \dodoi{10.1134/S1063773713020060}

\bibitem[{{Planck Collaboration} {et~al.}(2020){Planck Collaboration},
  {Aghanim}, {Akrami}, {Ashdown}, {Aumont}, {Baccigalupi}, {Ballardini},
  {Banday}, {Barreiro}, {Bartolo}, {Basak}, {Battye}, {Benabed}, {Bernard},
  {Bersanelli}, {Bielewicz}, {Bock}, {Bond}, {Borrill}, {Bouchet}, {Boulanger},
  {Bucher}, {Burigana}, {Butler}, {Calabrese}, {Cardoso}, {Carron},
  {Challinor}, {Chiang}, {Chluba}, {Colombo}, {Combet}, {Contreras}, {Crill},
  {Cuttaia}, {de Bernardis}, {de Zotti}, {Delabrouille}, {Delouis}, {Di
  Valentino}, {Diego}, {Dor{\'e}}, {Douspis}, {Ducout}, {Dupac}, {Dusini},
  {Efstathiou}, {Elsner}, {En{\ss}lin}, {Eriksen}, {Fantaye}, {Farhang},
  {Fergusson}, {Fernandez-Cobos}, {Finelli}, {Forastieri}, {Frailis},
  {Fraisse}, {Franceschi}, {Frolov}, {Galeotta}, {Galli}, {Ganga},
  {G{\'e}nova-Santos}, {Gerbino}, {Ghosh}, {Gonz{\'a}lez-Nuevo}, {G{\'o}rski},
  {Gratton}, {Gruppuso}, {Gudmundsson}, {Hamann}, {Handley}, {Hansen},
  {Herranz}, {Hildebrandt}, {Hivon}, {Huang}, {Jaffe}, {Jones}, {Karakci},
  {Keih{\"a}nen}, {Keskitalo}, {Kiiveri}, {Kim}, {Kisner}, {Knox},
  {Krachmalnicoff}, {Kunz}, {Kurki-Suonio}, {Lagache}, {Lamarre}, {Lasenby},
  {Lattanzi}, {Lawrence}, {Le Jeune}, {Lemos}, {Lesgourgues}, {Levrier},
  {Lewis}, {Liguori}, {Lilje}, {Lilley}, {Lindholm}, {L{\'o}pez-Caniego},
  {Lubin}, {Ma}, {Mac{\'\i}as-P{\'e}rez}, {Maggio}, {Maino}, {Mandolesi},
  {Mangilli}, {Marcos-Caballero}, {Maris}, {Martin}, {Martinelli},
  {Mart{\'\i}nez-Gonz{\'a}lez}, {Matarrese}, {Mauri}, {McEwen}, {Meinhold},
  {Melchiorri}, {Mennella}, {Migliaccio}, {Millea}, {Mitra},
  {Miville-Desch{\^e}nes}, {Molinari}, {Montier}, {Morgante}, {Moss}, {Natoli},
  {N{\o}rgaard-Nielsen}, {Pagano}, {Paoletti}, {Partridge}, {Patanchon},
  {Peiris}, {Perrotta}, {Pettorino}, {Piacentini}, {Polastri}, {Polenta},
  {Puget}, {Rachen}, {Reinecke}, {Remazeilles}, {Renzi}, {Rocha}, {Rosset},
  {Roudier}, {Rubi{\~n}o-Mart{\'\i}n}, {Ruiz-Granados}, {Salvati}, {Sandri},
  {Savelainen}, {Scott}, {Shellard}, {Sirignano}, {Sirri}, {Spencer},
  {Sunyaev}, {Suur-Uski}, {Tauber}, {Tavagnacco}, {Tenti}, {Toffolatti},
  {Tomasi}, {Trombetti}, {Valenziano}, {Valiviita}, {Van Tent}, {Vibert},
  {Vielva}, {Villa}, {Vittorio}, {Wandelt}, {Wehus}, {White}, {White},
  {Zacchei}, \& {Zonca}}]{planck2020}
{Planck Collaboration}, {Aghanim}, N., {Akrami}, Y., {et~al.} 2020, \aap, 641,
  A6, \dodoi{10.1051/0004-6361/201833910}

\bibitem[{{Porayko} {et~al.}(2018){Porayko}, {Zhu}, {Levin}, {Hui}, {Hobbs},
  {Grudskaya}, {Postnov}, {Bailes}, {Bhat}, {Coles}, {Dai}, {Dempsey}, {Keith},
  {Kerr}, {Kramer}, {Lasky}, {Manchester}, {Os{\l}owski}, {Parthasarathy},
  {Ravi}, {Reardon}, {Rosado}, {Russell}, {Shannon}, {Spiewak}, {van Straten},
  {Toomey}, {Wang}, {Wen}, {You}, \& {PPTA Collaboration}}]{Porayko2018}
{Porayko}, N.~K., {Zhu}, X., {Levin}, Y., {et~al.} 2018, \prd, 98, 102002,
  \dodoi{10.1103/PhysRevD.98.102002}

\bibitem[{{Sameie} {et~al.}(2020){Sameie}, {Yu}, {Sales}, {Vogelsberger}, \&
  {Zavala}}]{Sameie2020}
{Sameie}, O., {Yu}, H.-B., {Sales}, L.~V., {Vogelsberger}, M., \& {Zavala}, J.
  2020, \prl, 124, 141102, \dodoi{10.1103/PhysRevLett.124.141102}

\bibitem[{{Sandage}(1962)}]{sandage1962}
{Sandage}, A. 1962, \apj, 136, 319, \dodoi{10.1086/147385}

\bibitem[{{Schwab} {et~al.}(2016){Schwab}, {Rakich}, {Gong}, {Mahadevan},
  {Halverson}, {Roy}, {Terrien}, {Robertson}, {Hearty}, {Levi}, {Monson},
  {Wright}, {McElwain}, {Bender}, {Blake}, {St{\"u}rmer}, {Gurevich},
  {Chakraborty}, \& {Ramsey}}]{NEID_optical}
{Schwab}, C., {Rakich}, A., {Gong}, Q., {et~al.} 2016, in \procspie, Vol. 9908,
  Ground-based and Airborne Instrumentation for Astronomy VI, 99087H,
  \dodoi{10.1117/12.2234411}

\bibitem[{{Scolnic} {et~al.}(2018){Scolnic}, {Jones}, {Rest}, {Pan},
  {Chornock}, {Foley}, {Huber}, {Kessler}, {Narayan}, {Riess}, {Rodney},
  {Berger}, {Brout}, {Challis}, {Drout}, {Finkbeiner}, {Lunnan}, {Kirshner},
  {Sanders}, {Schlafly}, {Smartt}, {Stubbs}, {Tonry}, {Wood-Vasey}, {Foley},
  {Hand}, {Johnson}, {Burgett}, {Chambers}, {Draper}, {Hodapp}, {Kaiser},
  {Kudritzki}, {Magnier}, {Metcalfe}, {Bresolin}, {Gall}, {Kotak}, {McCrum}, \&
  {Smith}}]{scolnic2018}
{Scolnic}, D.~M., {Jones}, D.~O., {Rest}, A., {et~al.} 2018, \apj, 859, 101,
  \dodoi{10.3847/1538-4357/aab9bb}

\bibitem[{{Silverwood} \& {Easther}(2019)}]{Silverwood2019}
{Silverwood}, H., \& {Easther}, R. 2019, \pasa, 36, e038,
  \dodoi{10.1017/pasa.2019.25}

\bibitem[{Simon(2018)}]{Simon2018}
Simon, J.~D. 2018, The Astrophysical Journal, 863, 89,
  \dodoi{10.3847/1538-4357/aacdfb}

\bibitem[{Stankus {et~al.}(2020)Stankus, Nomerotski, Slo{\v{z}}ar, \&
  Vintskevich}]{stankus2020}
Stankus, P., Nomerotski, A., Slo{\v{z}}ar, A., \& Vintskevich, S. 2020, arXiv
  preprint arXiv:2010.09100

\bibitem[{Steinmetz {et~al.}(2020)Steinmetz, Matijevi{\v{c}}, Enke, Zwitter,
  Guiglion, McMillan, Kordopatis, Valentini, Chiappini, Casagrande, Wojno,
  Anguiano, Bienaym{\'{e}}, Bijaoui, Binney, Burton, Cass, de~Laverny, Fiegert,
  Freeman, Fulbright, Gibson, Gilmore, Grebel, Helmi, Kunder, Munari, Navarro,
  Parker, Ruchti, Recio-Blanco, Reid, Seabroke, Siviero, Siebert, Stupar,
  Watson, Williams, Wyse, Anders, Antoja, Birko, Bland-Hawthorn, Bossini,
  Garc{\'{\i}}a, Carrillo, Chaplin, Elsworth, Famaey, Gerhard, Jofre, Just,
  Mathur, Miglio, Minchev, Monari, Mosser, Ritter, Rodrigues, Scholz, Sharma,
  \& and}]{Steinmetz2020}
Steinmetz, M., Matijevi{\v{c}}, G., Enke, H., {et~al.} 2020, The Astronomical
  Journal, 160, 82, \dodoi{10.3847/1538-3881/ab9ab9}

\bibitem[{{Szentgyorgyi}(2017)}]{GCLEF}
{Szentgyorgyi}, A. 2017, in ESO Calibration Workshop: The Second Generation VLT
  Instruments and Friends, 46, \dodoi{10.5281/zenodo.887311}

\bibitem[{ten Brummelaar {et~al.}(2005)ten Brummelaar, McAlister, Ridgway,
  W.~G.~Bagnuolo, Turner, Sturmann, Sturmann, Berger, Ogden, Cadman, Hartkopf,
  Hopper, \& Shure}]{tenBrummelaar2005}
ten Brummelaar, T.~A., McAlister, H.~A., Ridgway, S.~T., {et~al.} 2005, The
  Astrophysical Journal, 628, 453, \dodoi{10.1086/430729}

\bibitem[{Tsai {et~al.}(2021{\natexlab{a}})Tsai, Eby, \&
  Safronova}]{Tsai:2021lly}
Tsai, Y.-D., Eby, J., \& Safronova, M.~S. 2021{\natexlab{a}}.
\newblock \doarXiv{2112.07674}

\bibitem[{Tsai {et~al.}(2021{\natexlab{b}})Tsai, Wu, Vagnozzi, \&
  Visinelli}]{Tsai:2021irw}
Tsai, Y.-D., Wu, Y., Vagnozzi, S., \& Visinelli, L. 2021{\natexlab{b}}.
\newblock \doarXiv{2107.04038}

\bibitem[{{Tulin} \& {Yu}(2018)}]{Tulin_Yu2018}
{Tulin}, S., \& {Yu}, H.-B. 2018, \physrep, 730, 1,
  \dodoi{10.1016/j.physrep.2017.11.004}

\bibitem[{{Vargya} {et~al.}(2021){Vargya}, {Sanderson}, {Sameie},
  {Boylan-Kolchin}, {Hopkins}, {Wetzel}, \& {Graus}}]{Vargya2021}
{Vargya}, D., {Sanderson}, R., {Sameie}, O., {et~al.} 2021, arXiv e-prints,
  arXiv:2104.14069.
\newblock \doarXiv{2104.14069}

\bibitem[{Vogt {et~al.}(1994)Vogt, Allen, Bigelow, Bresee, Brown, Cantrall,
  Conrad, Couture, Delaney, Epps, Hilyard, Hilyard, Horn, Jern, Kanto, Keane,
  Kibrick, Lewis, Osborne, Pardeilhan, Pfister, Ricketts, Robinson, Stover,
  Tucker, Ward, \& Wei}]{Vogt1994}
Vogt, S.~S., Allen, S.~L., Bigelow, B.~C., {et~al.} 1994, in Instrumentation in
  Astronomy {VIII}, ed. D.~L. Crawford \& E.~R. Craine ({SPIE}),
  \dodoi{10.1117/12.176725}

\bibitem[{{Wright} \& {Robertson}(2017)}]{WrightRobertson}
{Wright}, J.~T., \& {Robertson}, P. 2017, Research Notes of the American
  Astronomical Society, 1, 51, \dodoi{10.3847/2515-5172/aaa12e}

\bibitem[{Wyrzykowski {et~al.}(2016)Wyrzykowski, Kostrzewa-Rutkowska, Skowron,
  Rybicki, Mr{\'{o}}z, Koz{\l}owski, Udalski, Szyma{\'{n}}ski,
  Pietrzy{\'{n}}ski, Soszy{\'{n}}ski, Ulaczyk, Pietrukowicz, Poleski, Pawlak,
  I{\l}kiewicz, \& Rattenbury}]{Wyrzykowski2016}
Wyrzykowski, {\L}., Kostrzewa-Rutkowska, Z., Skowron, J., {et~al.} 2016,
  Monthly Notices of the Royal Astronomical Society, 458, 3012,
  \dodoi{10.1093/mnras/stw426}

\bibitem[{Zhang {et~al.}(2020)Zhang, England, Nomerotski, Svihra, Ferrante,
  Hockett, \& Sussman}]{Zhang2020}
Zhang, Y., England, D., Nomerotski, A., {et~al.} 2020, Phys. Rev. A, 101,
  053808, \dodoi{10.1103/PhysRevA.101.053808}

\bibitem[{Zhu {et~al.}(2020)Zhu, Colangelo, Chen, Korzh, Wong, Shaw, \&
  Berggren}]{Zhu2020}
Zhu, D., Colangelo, M., Chen, C., {et~al.} 2020, Nano Letters,
  \dodoi{10.1021/acs.nanolett.0c00985}

\end{thebibliography}

\end{document}